\documentclass[11pt]{article}
\usepackage{moriond,epsfig}
\usepackage{floatflt}

\def\Journal#1#2#3#4{{#1} {\bf #2}, #3 (#4)}

\def\PRD{{\em Phys. Rev.} D}

\def\ApJ{{\em Astrophys. J.}}
\def\MNRAS{{\em MNRAS}}

\def\be{\begin{equation}}
\def\ee{\end{equation}}
\def\bea{\begin{eqnarray}}
\def\eea{\end{eqnarray}}

\begin{document}
\vspace*{4cm}
\title{THE EFFECT OF SUPERNOVA ASYMMETRY ON COALESCENCE RATES OF
BINARY NEUTRON STARS}

\author{ K.A. POSTNOV, A.G. KURANOV }

\address{Sternberg Astronomical Institute, Universitetskij pr. 13,\\
119992 Moscow, Russia}

\maketitle\abstracts{We study the effect of the kick velocity -- neutron star spin 
alignment observed in young radio pulsars on the coalescence rate of binary neutron stars.
The effect is shown to be especially strong for large kick amplitudes and tight alignments, reducing
the galactic rate of binary neutron star coalescences up to an order of magnitude
with respect to the rates calculated for random kicks. The spin-kick correlation also 
leads to much narrower NS spin-orbit misalignments compared to random kicks.}

\section{Introduction}
Kick velocity imparted to a newborn neutron star is known to be
an important phenomenological parameter of core collapse supernovae. 
The origin of the kicks remains unclear (see, for example, \cite{Lai} 
and references therein). Recently, new important observational results
appeared suggesting possible NS spin-kick alignment. Tight spin-kick alignment 
follows from measurements of radio pulsar polarization \cite{Johnston_ea}, 
as well as from X-ray observations
of pulsar wind nebulae around young pulsars \cite{Helfand,Kargaltsev}. 
Implications of these observations to the formation of double 
pulsars were discussed by Wang et al. \cite{Wang_ea06}. 
Here we explore the effect of NS spin -- kick correlation 
on the formation and galactic coalescence rate of double neutron stars (DNS) which
are primary targets for modern gravitational wave detectors. We show that the tighter alignement,
the smaller is the DNS merging rate with respect to models with random kick orientation. The effect 
is especially important for large kick amplitudes ($\sim 400$ km/s). 

\section{Effect on the binary neutron star coalescence rates}

The effect of NS kick velocity on merging rates of compact binaries was
studied earlier (e.g. \cite{Lipunov_ea97}).  
The observed tight NS spin -- kick alignment may have important implications to 
the formation and evolution of binary compact stars (see especially earlier
paper by Kalogera \cite{Kalogera}).
Let us consider the standard evolutionary scenario leading to formation of 
binary NS from a massive binary system 
(see \cite{PYu} for discussion and references) focusing
on the effect of the NS kick velocity. 
We shall assume that the kick velocity vector 
is confined within a cone which is coaxial with the progenitor's rotation axis and 
characterized by angle $\theta<\pi/2$. We shall consider only central kicks thus ignoring
theoretically feasible off-center kicks affecting the NS spin~\cite{Spruit,Wang_ea07}. 
The value of the kick velocity is assumed to obey 
the Maxellian distribution $f(v)\sim v^2 \exp(-(v/v_0)^2)$, as suggested by pulsar
proper motion measurements \cite{Hobbs}. The velocity $v_0$ varied from 0 to 400 km/s. 

The rotational axes of both components are assumed to
be aligned with the orbital angular momentum before the primary collapses to form first NS. 
The SN explosion is treated in a standard way as instantaneous 
loss of mass of the exploding star.  
The effect of the kick on the post-explosion binary orbital parameters is treated from the
point of view of energy-momentum conservation in two point-mass body problem 
(see e.g. in \cite{Kalogera,Grishchuk}). The first SN explosion most likely occurs when the binary orbit is circular (unless the initial binary is very wide so that tidal circularization is ineffective), while the second explosion can
happen before the orbit has been tidally circularized; in the latter case we choose the
position of the star in the orbit distributed according to Kepler's 2d law. 

\begin{figure}
\psfig{figure=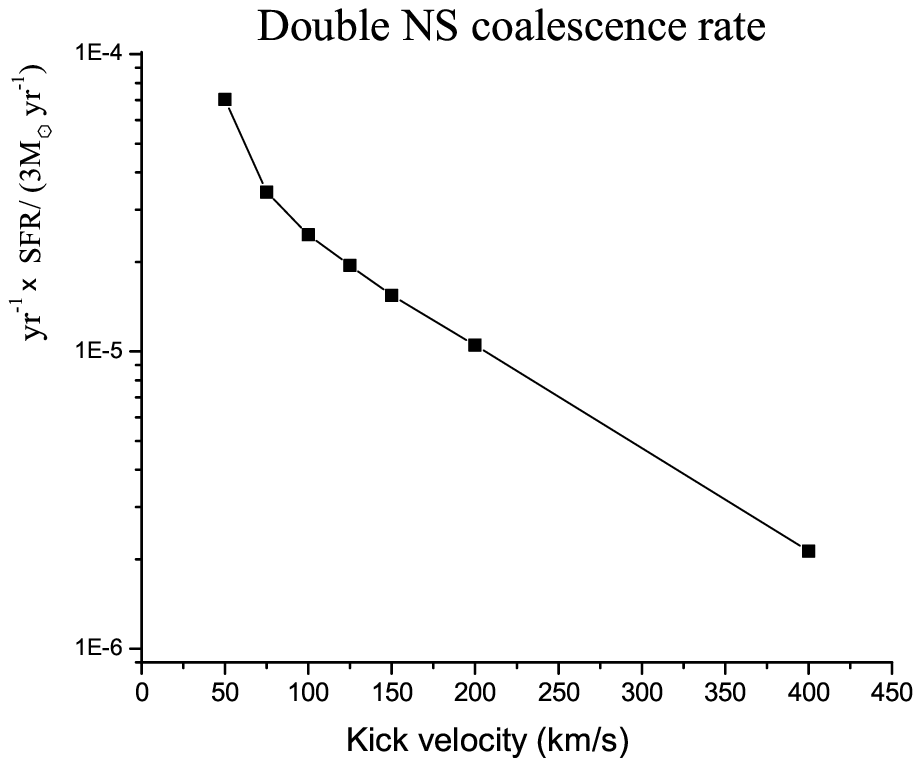,height=6cm,width=0.4\textwidth}
\hfill
\psfig{figure=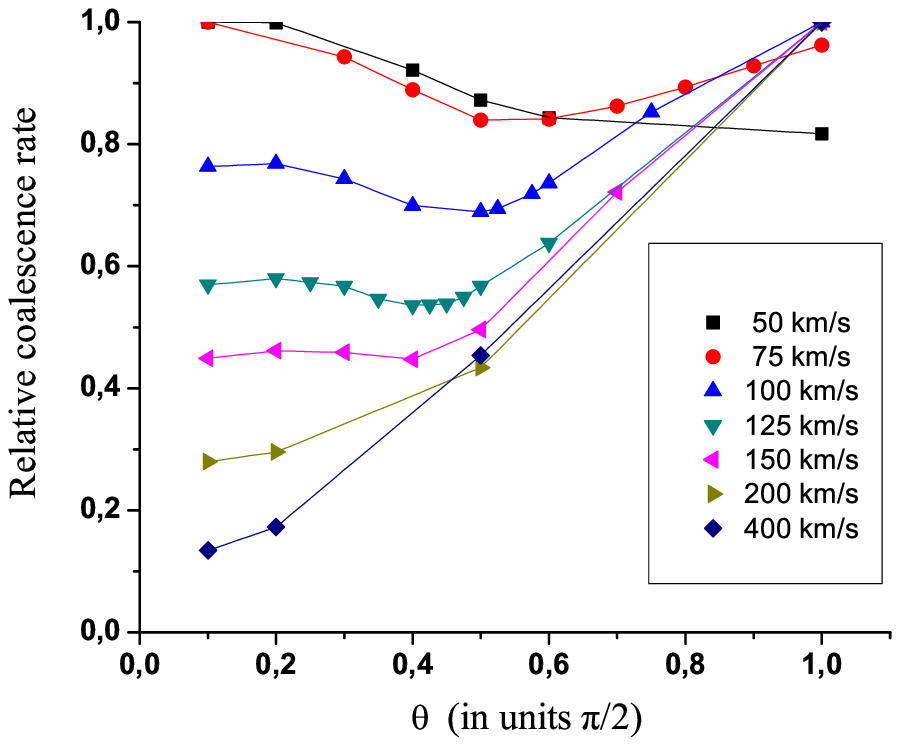,height=6cm,width=0.4\textwidth}
\parbox[7]{0.47\textwidth}{\caption{Galactic coalescence rate of DNS vs. kick parameter $v_0$ (random kicks). An almost exponential decay with $v_0$ is seen for $v_0>100$ km/s}}
\hfill
\parbox[7]{0.47\textwidth}{\caption{Relative change of DNS merging rate for NS spin-kick correlation}}
\end{figure}

We use the population synthesis method to calculate the expected coalescence rate 
of DNS (see \cite{Lipunov_ea97,PYu} and references therein). The standard assumptions about
binary evolution have been made: Salpeter's mass function for the primary's mass, $f(M_1)\sim M^{-2.35}$, 
flat initial mass ratio ($q=M_2/M_1<1$) distribution $f(q)=const$, initial semi-major axes distribution
in the form $d\log a=const$. The common envelope phase is treated in the standard way  \cite{PYu}  
with the efficiency $\alpha_{CE}=0.5$. 
The calculations were normalized to the galactic star formation rate 
$3M_\odot$ per year, with binary fraction 50\%. We also have carefully taken into account rotational evolution 
of magnetized compact stars, as described in \cite{Lipunov,LPP96}, 
assuming no magnetic field decay. The galactic DNS merging rate 
is shown in Fig. 1 as a function of the kick parameter $v_0$ 
and assuming random central kicks. Note an almost exponential decay 
of the rate with $v_0$ for $v_0>100$ km/s. Fig. 2 shows the relative change 
in the DNS merging rate when we allow for NS spin-kick alignment with different values of the
confinement angle $\theta$. It is seen that tight alignment generally reduces the DNS merging rate, 
with the effect being especially strong for large kick velocity amplitudes. Such a reducing relative to 
calculations with random kicks is in fact expected, because the NS spin -- kick correlation 
excludes kicks in the orbital plane which, if directed opposite to the orbital velocity, 
can additionally bind the post-explosion binary system.  

\section{Neutron star spin -- orbit misalignment}

\begin{figure}
\psfig{figure=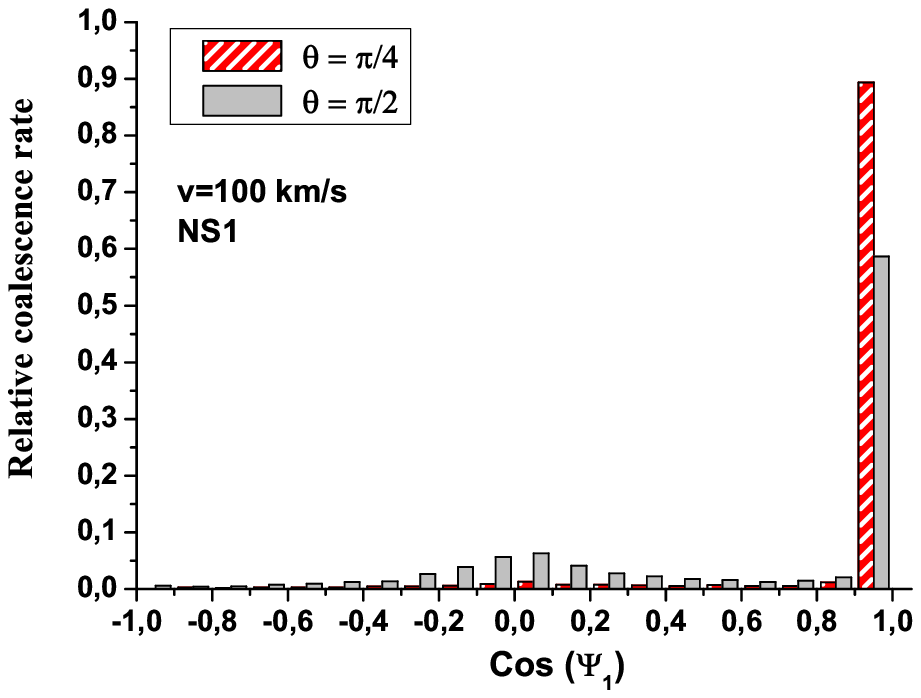,height=4.3cm,width=0.24\textwidth}
\psfig{figure=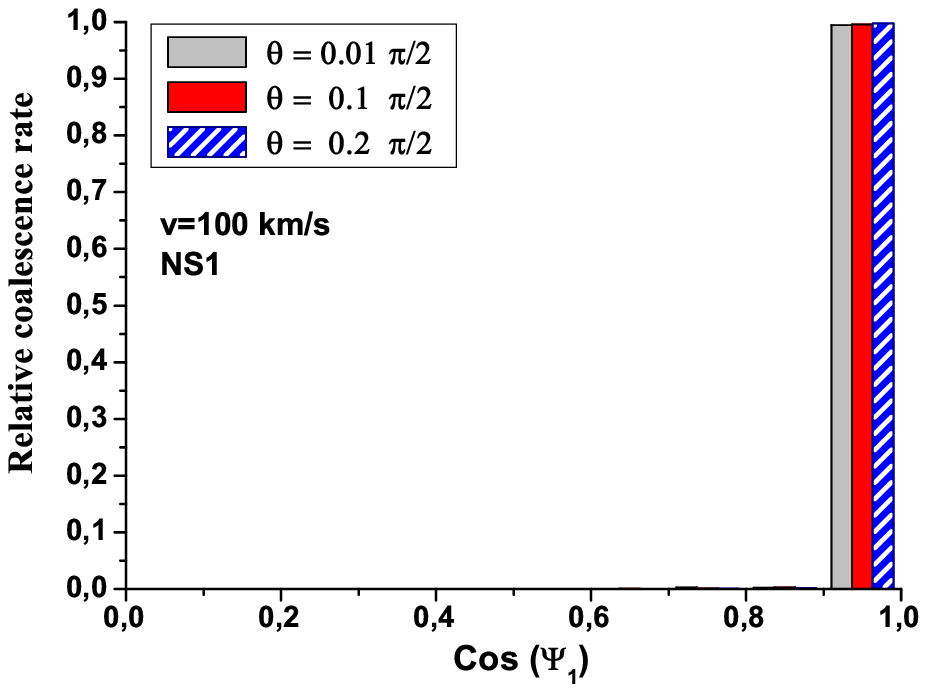,height=4.3cm,width=0.24\textwidth}
\hfill
\psfig{figure=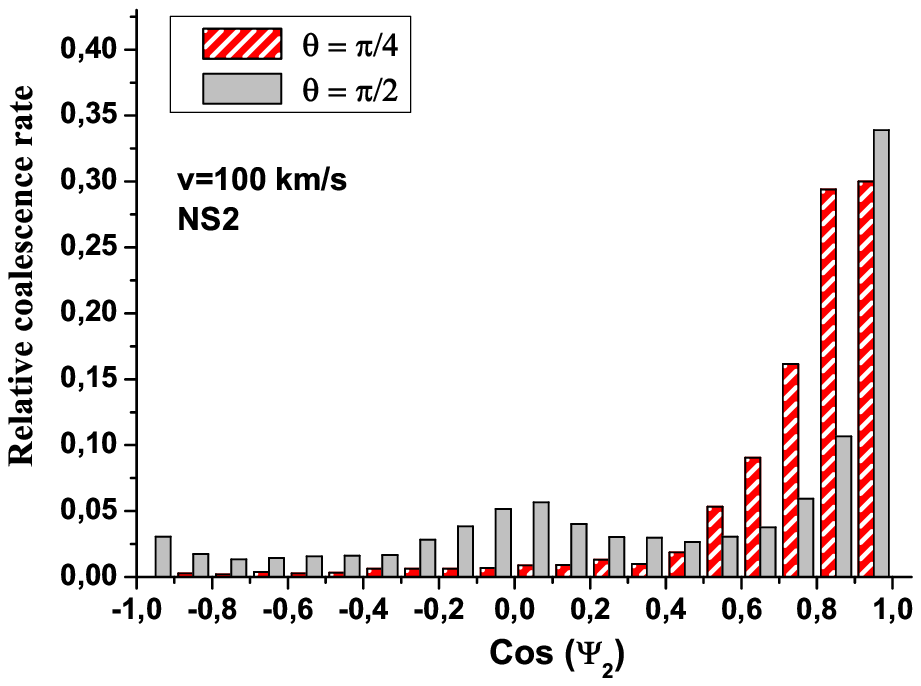,height=4.3cm,width=0.24\textwidth}
\psfig{figure=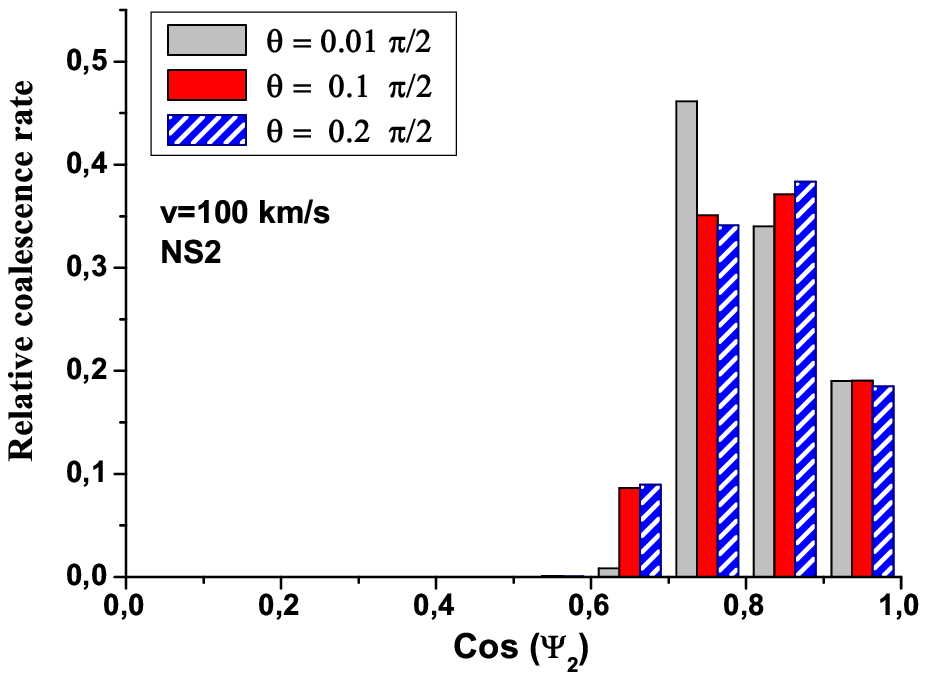,height=4.3cm,width=0.24\textwidth}
% \rule{\textwidth}{0.2mm}
\psfig{figure=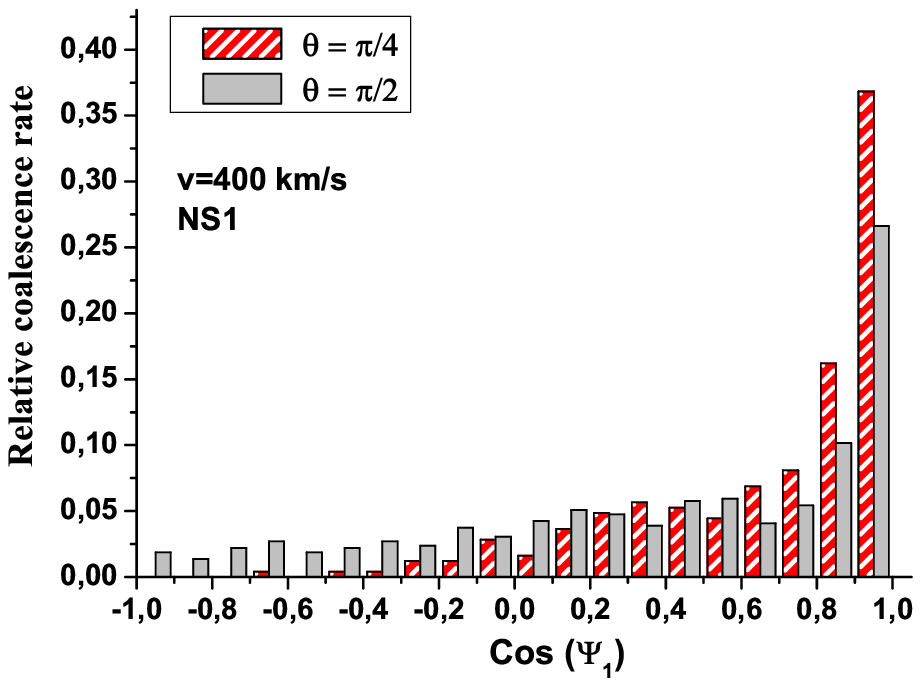,height=4.3cm,width=0.24\textwidth}
\psfig{figure=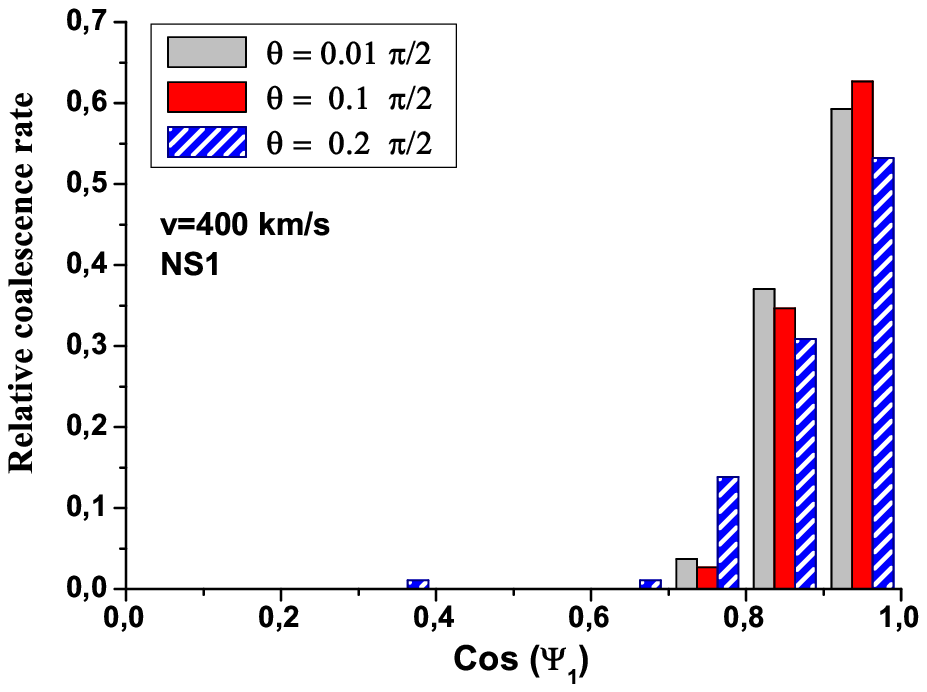,height=4.3cm,width=0.24\textwidth}
\hfill
\psfig{figure=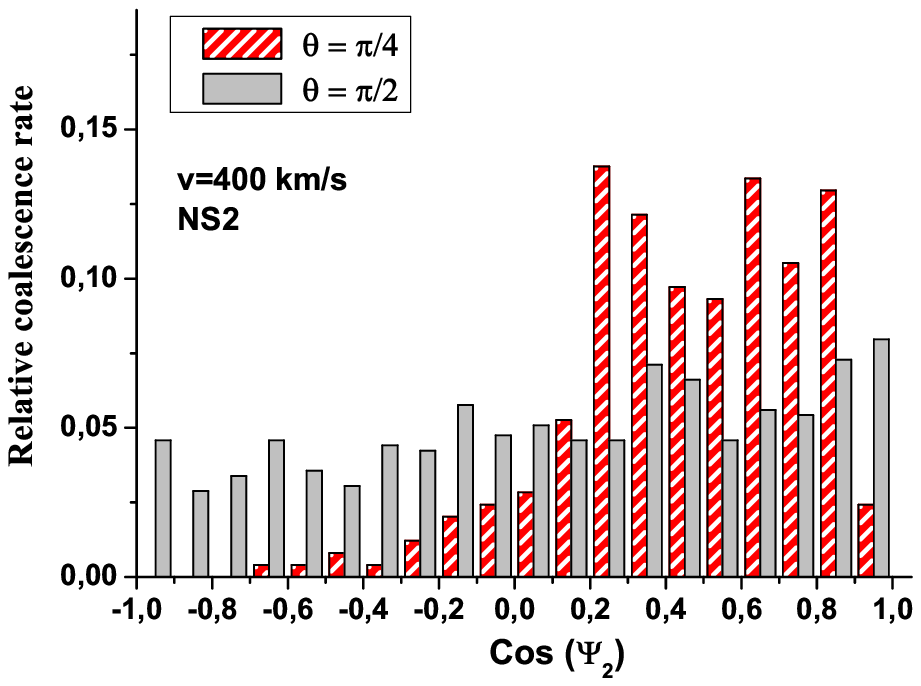,height=4.3cm,width=0.24\textwidth}
\psfig{figure=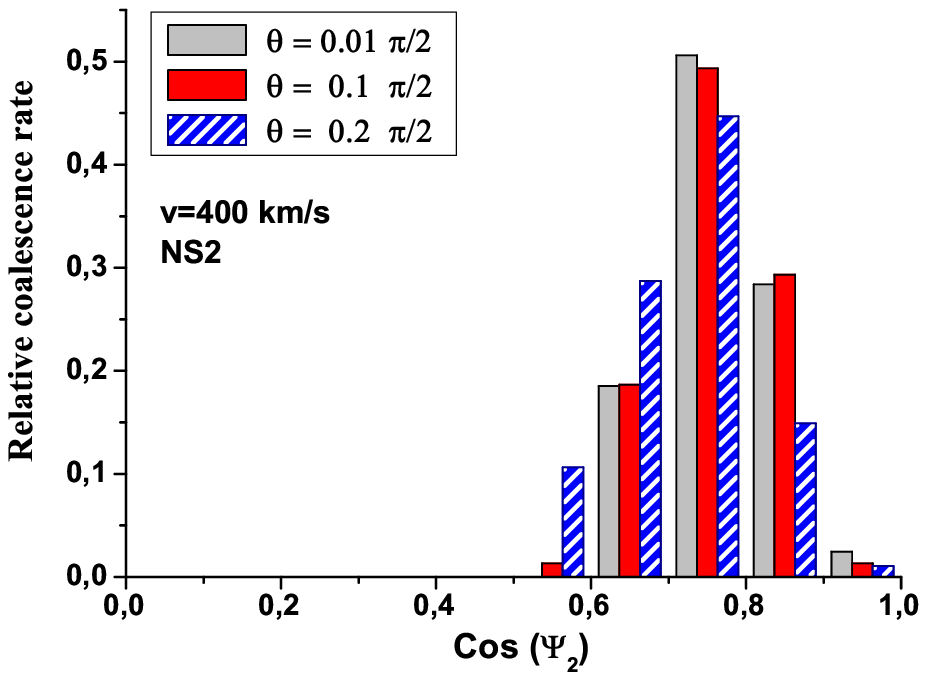,height=4.3cm,width=0.24\textwidth}
\caption{NS spin-orbit misalignment $\cos\Psi$ in coalescing DNS for $v_0=100$ km/s (upper row) and 400 km/s (bottom row) and different NS spin-kick alignment angles $\theta_0$. NS1: spin aligned with orbit
before SN explosion; NS2: spin aligned with original binary's orbital angular momentum}
% % \parbox[7]{0.47\textwidth}{\caption{Galactic coalescence rate of DNS vs. kick parameter $v_0$ (random kicks)}}
% \hfill
% \parbox[7]{0.47\textwidth}{\caption{Relative change of DNS merging rate for NS spin-kick alignment}}
\end{figure}

There is another observational 
consequence of the kick in DNS systems: NS spin -- orbit misalignment, which can be tested by geodetic precession measurements in binary pulsars \cite{Bailes}. Such a misalignment 
is potentially very interesting for GW studies \cite{Apostolatos}. 
After SN explosion in a binary system, additional kick imparted to newborn NS 
results, in general, in a misalignment between
the new orbital angular momentum and the NS (and the secondary component's) 
spin vector characterized by angle $\Psi_1$. After the second SN explosion in the system, there are several possibilities for the NS spin-orbit misalignment.  

1) In sufficiently wide binaries, when
tidal interactions between the components are inefficient, the orientation of the NS1 spin
and the secondary's spin vector may remain unchaged until the second SN explosion, after which
the orbital angular momentum vector changes again due to NS2 kick. So in this case we would 
expect two coaxial NS with spins misaligned by angle $\Psi_2$ with orbital angular momentum. 
However, such binaries, unless very eccentric, may be too wide
to coalesce over the Hubble time. 

2) In close binaries, tidal interactions tend to rapidly align angular momentum 
vector of the normal star with the orbital angular momentum. To spin-up the 
NS rotation up to observed ms periods (in binary ms pulsars), a modest amount of matter ($\sim 0.1 M_\odot$) should
be accreted by NS. This amount is sufficient to align the NS rotation with the orbital angular momentum.
So if NS1 accretes matter before the second SN explosion,  
both NS1 and the secondary component's spins are most likely aligned with orbital
angular momentum (see discussion in \cite{Wang_ea06}). Note that the NS1 spin tends to 
align with the orbital angular momentum even if NS1 does not accrete matter 
but spins-down by the propeller mechanism before the second SN explosion, since in that case very strong currents must flow through its polar cap and the alignment torques is as strong as during accretion.
So the NS1 remains misaligned prior to the second SN explosion only in rare cases where the secondary collapses shortly after the first SN in the binary.
If both NS1 and secondary were aligned with orbital angular momentum prior to the second SN explosion, 
both neutron stars will be equally misaligned with orbital angular momentum, $\Psi_1=\Psi_2$.

In our population synthesis simulations we take into account the discussed spin alignment effects.
In Fig. 3 we show the calculated distribution between the NS1 and NS2 spins and
orbital angular momentum in coalescing DNS systems (angles $\Psi_1$ and $\Psi_2$, respectively)
assuming spin-orbit alignment (angle $\Psi_1$) and conservation of the secondary's angular momentum
(angle $\Psi_2$). Clearly, the real distribution must be intermediate between the two, depending
on the degree of misalignment of the secondary's angular momentum prior to the collapse. 
It is seen that the misalignment angles 
can be very different (and even with negative cosines) for random or loosely constrained 
($\theta \sim \pi/2$) kicks, while tight spin-kick 
alignment ($\theta\ll \pi/2$) results in much narrow distributions (see also \cite{Kalogera}). 
The mean misalignement angles $\Psi$
are presented in Table 1.

\begin{table}
\caption{Mean NS spin-orbit misalignment $\Psi$ (in units $\pi/2$)}
\bigskip
\begin{center}
%\begin{tabular}{|r|c|c|c|c|c|}
\begin{tabular}{|c|ccccc|ccccc|}
\hline
Kick $v_0$&\multicolumn{5}{|c}{NS1 ($\Psi_1$) }&\multicolumn{5}{|c|}{NS2 ($\Psi_2$)}  \\
\cline{2-11}
(km/s) &\multicolumn{10}{|c|}{Kick confinement angle $\theta$ (in units $\pi/2$)}\\
% &  Максимальная величина полярного угла(в единицах $\pi/2$) \\
\cline{2-11}
&\bf{ 0.01} & \bf{0.1} & \bf{0.2} & \bf{0.5}& \bf{1.0} & \bf{0.01} & \bf{0.1} & \bf{0.2} & \bf{0.5}& \bf{1.0} \\
% \hline
% NS 1\\
%\multicolumn{1}{r}{Star 1}&\\
\hline
\bf{ 50}  &   0.061  &   0.060 &  0.059 &  0.073 &  0.212&   0.307  &  0.310  &  0.316 &  0.308  & 0.345 \\
\bf{100}  &   0.110  &   0.110 &  0.109 &  0.186 &  0.444&   0.378  &  0.381  &  0.381 &  0.423  & 0.633\\
\bf{200}  &   0.192  &   0.195 &  0.207 &  0.337 &  0.535&   0.417  &  0.419  &  0.442 &  0.575  & 0.813\\
\bf{400}  &   0.257  &   0.262 &  0.291 &  0.451 &  0.670&   0.442  &  0.447  &  0.481 &  0.670  & 0.909\\
\hline
\end{tabular}
\end{center}
\end{table}

\section{Conclusions}

We have shown that the spin-velocity correlation observed in radio pulsars,
suggesting NS spin-kick velocity alignment, may have very important implications
to GW studies. First, the tight alignment reduces the galactic rate 
of double neutron star coalescences (especially for large kicks 300-400 km/s -- up to ten times)
relative to models with random kicks. Second, the spin-kick correlation results in 
specific distribution of NS spin -- orbit misalignments, which can be tested 
by analysing GW signals from DNS mergings.  

\section*{Acknowledgments}
KAP acknowledges the financial support from the Meeting Organizers 
and RFBR grant 07-02-08065z.

\end{document}